\documentstyle[amstex,preprint,psfig,aps]{revtex}

\begin{document}
\title{Observation of neutral current charm production in $\nu _{\mu }Fe$
scattering at the Tevatron}
\author{ A.~Alton, T.~Adams, T.~Bolton, J.~Goldman, M.~Goncharov}
\address{Kansas State University, Manhattan, KS, 66506}
\author{R.~A.~Johnson, N. Suwonjandee, M. Vakili}
\address{University of Cincinnati, Cincinnati, OH, 45221}
\author{J.~Conrad, B.~T.~Fleming, J.~Formaggio, 
J.~H.~Kim,\footnote[4]{Present 
Address: University of California, Irvine, CA, 92697}
S.~Koutsoliotas,\footnote[3]{Present Address:
Bucknell University, Lewisburg, PA, 17837}  C.~McNulty, 
A.~Romosan,\footnote[5]{Present Address: University of California, Berkeley, 
CA, 94720} M.~H.~Shaevitz, P.~Spentzouris,\footnote[6]{Present Address: Fermi 
National Laboratory, Batavia, IL, 60510} E.~G.~Stern, 
A.~Vaitaitis, E.~D.~Zimmerman}
\address{Columbia University, New York, NY, 10027}
\author{R.~H.~Bernstein, L.~Bugel, M.~J.~Lamm, W.~Marsh, 
P.~Nienaber,\footnote[7]{Present Address: Marquette University, Milwaukee, 
WI, 53201} J.~Yu}
\address{Fermi National Accelerator Laboratory, Batavia, IL, 60510}
\author{L.~de~Barbaro, D.~Buchholz, H.~Schellman, G.~P.~Zeller}
\address{Northwestern University, Evanston, IL, 60208}
\author{J.~Brau, R.~B.~Drucker, R.~Frey, D.~Mason}
\address{University of Oregon, Eugene, OR, 97403}
\author{J. McDonald, D.~Naples} 
\address{University of Pittsburgh, Pittsburgh, PA, 15260}
\author{S.~Avvakumov, P.~de~Barbaro, A.~Bodek, H.~Budd, 
D.~A.~Harris,\footnotemark[6]
K.~S.~McFarland, W.~K.~Sakumoto, U.~K.~Yang}
\address{University of Rochester, Rochester, NY, 14627}
\author{Version 1.0.0}
\date{\today}
\maketitle

\begin{abstract}
We report on the first observation of open charm production in neutral
current deep inelastic neutrino scattering as seen in the NuTeV\ detector at
Fermilab. The production rate is shown to be consistent with a pure gluon-$%
Z^{0}$ boson production model, and the observed level of charm production is
used to determine the effective charm mass. As part of our analysis, we also
obtain a new measurement for the proton-nucleon charm production cross
section at $\sqrt{s}=38.8$ GeV.
\end{abstract}

\section{ Introduction}

Measurements of charged current (CC) charm production in deep inelastic
(DIS) neutrino and anti-neutrino scattering have proven to be an excellent
source of information on the structure of the nucleon, the dynamics of heavy
quark production, and the values of several fundamental parameters of the
Standard Model (SM) of particle physics\cite{JC-MS-TB rmp}. The only evidence
to date of neutral current (NC) charm production in $\nu _{\mu }N$ or $\bar{%
\nu}_{\mu }N$ scattering is an unconfirmed observation of $J/\psi $
production\cite{bib:CDHS,todd prd}. \ Using a new Sign Selected Quadrupole
Train\ (SSQT) beam with the high energies of the Fermilab Tevatron, NuTeV is
able to perform a sensitive search for NC charm production through
detection of events with wrong sign muon (WSM) final states. This occurs
whenever an interaction produces a single muon and the muon has
the opposite lepton number as the neutrino beam. The SSQT produced event
ratios of $\bar{\nu}_{\mu }/\nu _{\mu }$ in neutrino and a $\nu _{\mu }/\bar{%
\nu}_{\mu }$ anti-neutrino mode of $0.8\times 10^{-3}$ and $4.8\times 10^{-3}
$, respectively, making possible the WSM identification.

In SM NC interactions, charm quarks must be produced in pairs. \ As is the
case for CC charm production, considerable suppression of the $\nu _{\mu
}N\rightarrow \nu _{\mu }c\bar{c}X$ rate occurs due to the non-zero charm
quark mass $m_{c}$. This measured suppression can be used to experimentally
determine $m_{c}$. If $m_{c}$ is a fundamental parameter of the SM, then it
should have the same value, up to possible quantum chromodynamics (QCD)
corrections, in NC and CC $\nu _{\mu }N$ charm production, and in other
physical processes such as the photo-production of charm and the spectrum of
charmonium.

\begin{figure}[tbp]
  \centerline{\psfig{figure=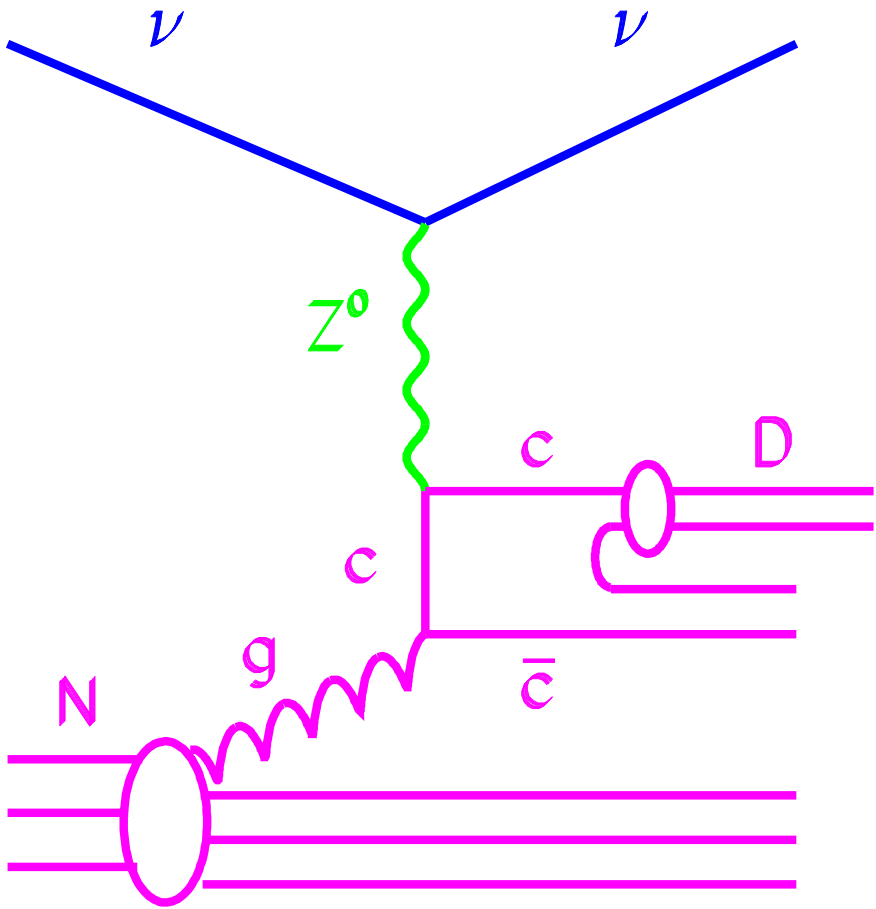,width=17cm}}
\caption{{}Feynman diagram for boson-gluon fusion}
\label{fig:bgf}
\end{figure}

In the fixed-flavor (FF) implementation of QCD\cite{bib:GGR}, NC charm pair
production can be attributed completely to hard scatters between the the
exchanged virtual $Z^{0}$ boson and a gluon in the nucleon sea; this
process, known as boson-gluon fusion, is illustrated in Fig. \ref{fig:bgf}.
Other implementations of QCD\cite{ACOT,MRST} envision an intrinsic charm
quark parton distribution function (PDF) $c\left( x,Q^{2}\right) $, that
depends on the Bjorken scaling variable $x$ and the absolute value of the
squared momentum transfer $Q^{2}$. In addition, some have suggested 
\cite{brodsky,ingelman,Navarra} 
that non-perturbative QCD effects may produce an unusually
large $c\left( x,Q^{2}\right) $, particularly at high $x$. In FF QCD, $%
c\left( x,Q^{2}\right) \simeq 0$ over the $Q^{2}$ range probed by NuTeV. The
validity of this assumption can be tested with the data.

\section{Experimental Apparatus and Beam}

\begin{figure}[tbp]
  \centerline{\psfig{figure=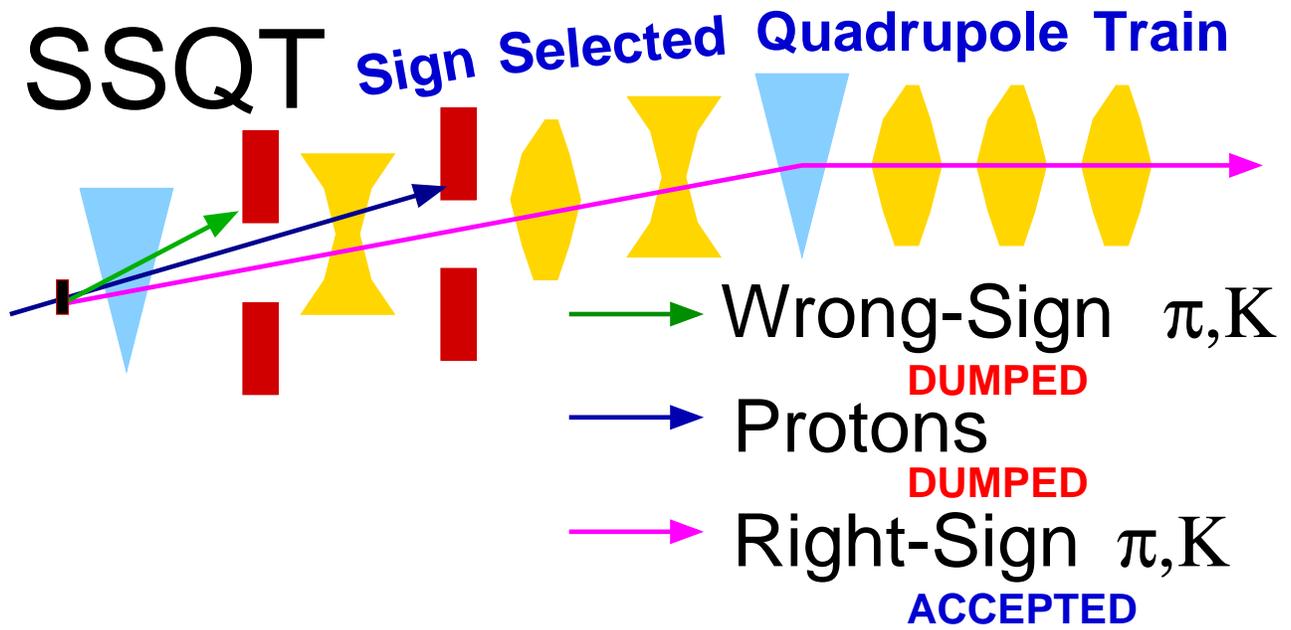,width=17cm}}
\caption{{}Schematic of the SSQT beamline}
\label{fig:ssqt}
\end{figure}

The NuTeV (Fermilab-E815) neutrino experiment collected data during 1996-97
with the refurbished Lab E neutrino detector and a newly installed
Sign-Selected Quadrupole Train(SSQT) neutrino beamline. Figure \ref{fig:ssqt}
illustrates the sign-selection optics employed by the SSQT to pick the
charge of secondary pions and kaons which determine whether $\nu _{\mu }$ or 
$\bar{\nu}_{\mu }$ are predominantly produced. During NuTeV's run the
primary production target received $1.13\times 10^{18}$ and $1.41\times
10^{18}$ protons-on-target in neutrino and anti-neutrino modes, respectively.

The Lab E detector, described in detail elsewhere\cite{bib:nim}, consists of
two major parts, a target calorimeter and an iron toroid spectrometer. The
target calorimeter contains 690 tons of steel sampled at 10 cm intervals by
84 3 m $\times $ 3 m scintillator counters and at 20 cm intervals by 42 3 m $%
\times $ 3 m drift chambers. The toroid spectrometer consists of four stations
of drift chambers separated by iron toroid magnets. Precision hadron and
muon calibration beams monitored the calorimeter and spectrometer
performance throughout the course of data taking. The calorimeter achieves a
sampling-dominated hadronic energy resolution of $\sigma
_{E_{HAD}}/E_{HAD}=2.4\%\oplus 87\%/\sqrt{E_{HAD}}$ and an absolute scale
uncertainty of $\delta E_{HAD}/E_{HAD}=0.5\%$. 
The spectrometer's muon energy resolution (dominated by multiple
Coulomb scattering) is $\sigma _{E_{\mu
}}/E_{\mu }=$ $11\%$ and the muon momentum scale is known to $\delta E_{\mu
}/E_{\mu }=1.0\%$. With the selection criteria used in this analysis, the
muon charge mis-identification probability in the spectrometer is $2\times
10^{-5}$.

\section{Analysis Procedure}

Much of the analysis procedure follows that used for a search for flavor
changing neutral currents using the same data set\cite{fcnc drew}, and
further details may be obtained from the article describing that analysis
and in Ref. \ref{drew thesis}.

\subsection{Introduction and Data Selection \label{sec:cut}}

The analysis technique consists of comparing the visible inelasticity, $%
y_{vis}=E_{HAD}/\left( E_{HAD}+E_{\mu }\right) $, measured in the $\nu _{\mu
} $ and $\bar{\nu}_{\mu }$ wrong sign muon (WSM) data samples to a Monte
Carlo (MC) simulation containing all known conventional WSM sources and a
possible NC charm signal. The NC charm signal peaks at large values of $\
y_{vis}$ because the decay muon from the heavy flavor hadron is usually much
less energetic than the hadron shower produced in the NC interaction. The
largest background, beam impurities, is concentrated at low $y_{vis}$ in $%
\nu _{\mu } $ mode due to the characteristic $\left( 1-y\right) ^{2}$
behavior of interactions of the $\bar{\nu}_{\mu }$ wrong-flavored beam
background.

Events in the WSM sample must satisfy a number of selection criteria
(``cuts''). The fiducial volume cut requires that event vertices be
reconstructed at least 25 cm-Fe (cm of iron) from the outer edges of the
detector in the transverse directions, at least 35 cm-Fe downstream of the
upstream face of the detector, and at least 200 cm-Fe upstream of the
toroid. \ Events must contain a hadronic energy of at least 10 GeV
(increased to 50 GeV for the final NC charm fit), and exactly one track (the
muon) must be found. The muon is required to be well-reconstructed and to
pass within the understood regions of the toroid's magnetic field. The
muon's energy must be between 10 and 150 GeV, and its charge must be
consistent with having the opposite lepton number as the primary beam
component. Requiring that the muon energy reconstructed in different
longitudinal sections of the toroid agree within 25\% of the value measured
using the full toroid reduces charge mis-identification backgrounds to the $%
2\times 10^{-5}$ level. This latter number has been verified using the muon
calibration beam.

\subsection{Source and Background Simulations \label{sec:ncc_mc}}

Conventional WSM sources arise from beam impurities, right-flavor CC events
where the charge of the muon is mis-reconstructed, CC and NC events where a $%
\pi $ or $K$ meson decays in the hadron shower, and CC charm production
where the primary muon is not reconstructed or the charm quark is produced
via a $\nu _{e}$ interaction. \ Table \ref{tab:fracs} lists the fractional
contribution of each background component. The relatively large beam
impurity background consists of contributions from hadrons (including charm)
that decay before the sign-selecting dipoles in the SSQT, neutral kaon
decays, muon decays, decay of hadrons produced by secondary interactions in
the SSQT (``scraping''), and from decay of wrong-sign pions produced in kaon
decays. Table \ref{tab:ws50} summarizes the relative contributions of each
beam source. For this analysis the beam sources can be further tuned using
WSM data in $\bar{\nu}_{\mu }$ mode. This procedure, which yields in passing a
new measurement of $\sigma \left( pN\rightarrow c\bar{c}X\right) $ at $\sqrt{%
s}=38.8$ GeV, is detailed in the Appendix.

After impurities, the next largest WSM source comes from CC production of
charm in which the charm quark decays semi-muonically, and its decay muon is
detected in the spectrometer. The primary lepton is either an electron,
which is lost in the hadrons shower, 
or a muon which exits from or ranges out in the calorimeter. The $\nu _{e}$
beam fraction is $1.9(1.3)\%$ in $\nu $ ($\bar{\nu}$) mode, and $22\%$ of
the CC charm events which pass WSM cuts originate from a $\nu _{e}$.\ 

\begin{figure}[tbp]
  \centerline{\psfig{figure=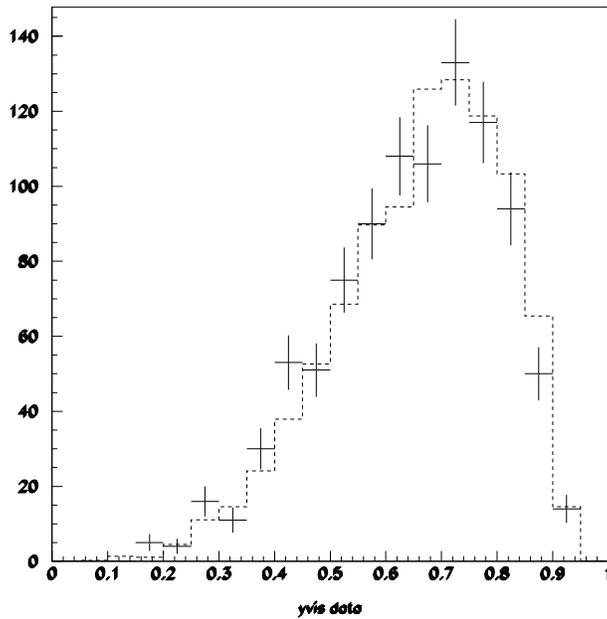,width=8cm}}
\caption{{}Comparison of data to MC of $E_{had}/(E_{had}+E_{\protect\mu 2})$
for dimuon events where both muons are toroid-analyzed.}
\label{fig:ditest}
\end{figure}

Charged current charm production produces a broad peak at high $y_{vis}$
that must be handled with care. The CC charm background is simulated using a
leading-order QCD charm production model with production, fragmentation, and
charm decay parameters tuned on neutrino dimuon data collected by NuTeV\cite
{bib:max} and a previous experiment using the same detector\cite{bib:baz}.
Overall normalization of the source is obtained from the measured
charm-to-total CC cross section ratio and the single muon right-sign data
sample. Simulated dimuon events are passed through the full GEANT simulation
of the detector. \ Figure \ref{fig:ditest} provides a check of the modelling
of this source through a comparison of the distribution of $y_{vis}^{\prime
}=$ $E_{HAD}/\left( E_{HAD}+E_{\mu 2}\right) $, where $E_{\mu 2}$ is the
energy of the WSM in the event, between data and MC for dimuon events in
which both muons are reconstructed by the spectrometer. \ This distribution
should closely mimic the expected background to the $y_{vis}$ distribution
in the WSM sample. A $\chi ^{2}$ comparison test between data and model
yields a value of 19 for 17 degrees of freedom.

Finally, a NC $c\bar{c}$ event produces a WSM when the $c\left( \bar{c}%
\right) $-quark decays semi-muonically in $\nu _{\mu }\left( \bar{\nu}_{\mu
}\right) $ mode. To compute effects of fragmentation, heavy quark decay,
acceptance, and resolution, production\ is simulated with a $Z^{0}$-gluon
fusion model \cite{bib:GGR} with charm mass parameter\footnote{%
The choice of a low input MC mass allows for consistent re-weighting of the
cross section over a wide range in the final fit procedure.} $m_{c}=0.5$ GeV%
/c$^{2}$ and the GRV94-HO\cite{bib:grv94} PDF set. No corrections for the
nuclear environment are applied, but possible effects are considered in the
systematic error. The NC\ charm quarks are fragmented and decayed using
procedures adapted from the CC charm modelling, and the resulting WSM events
are then simulated with the full detector MC and processed with the data
reconstruction code.

\section{Results and Interpretation}

\subsection{Fits to Data}

\begin{figure}[tbp]
  \centerline{\psfig{figure=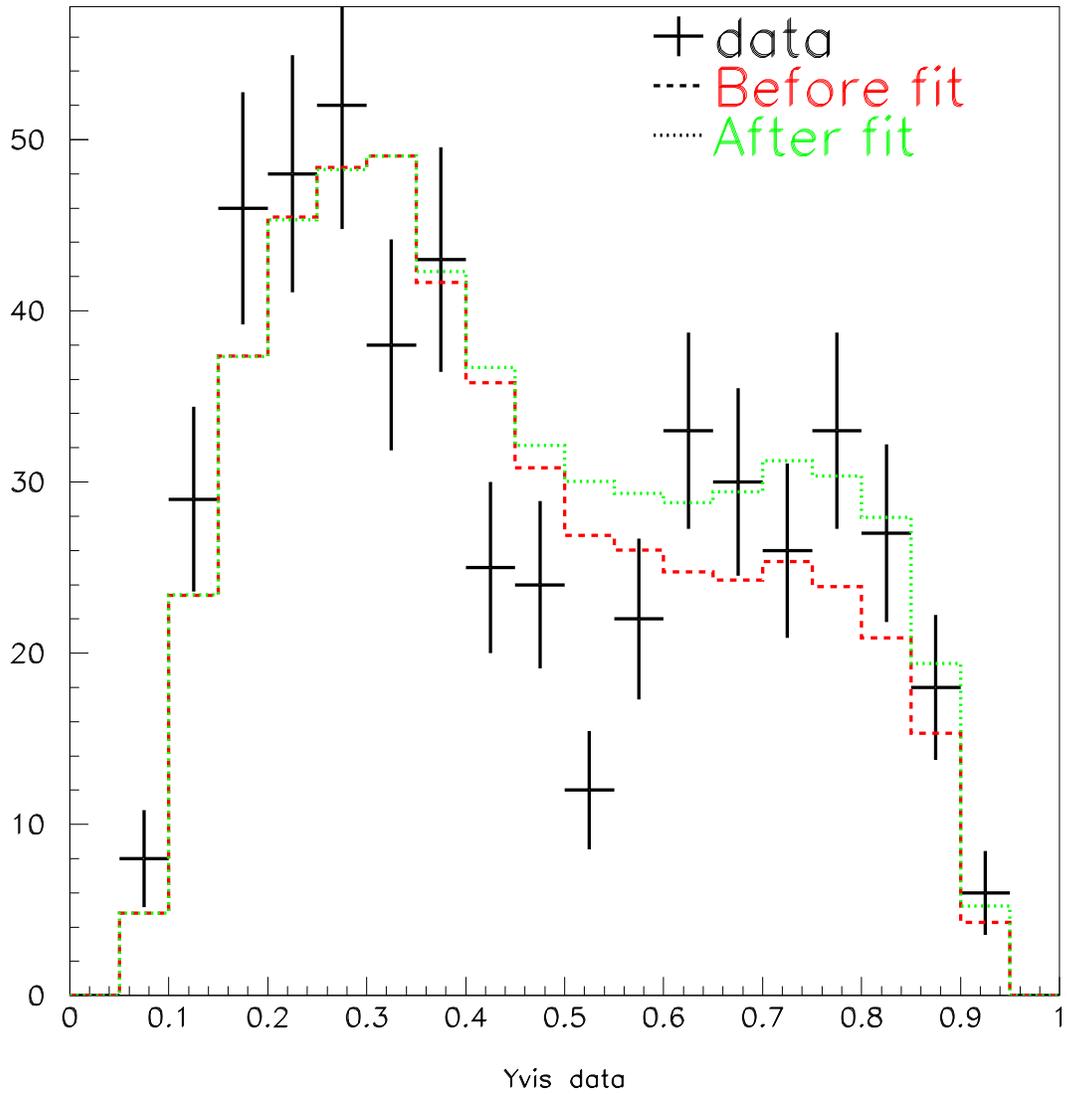,width=\textwidth}}
\caption{{}Distribution of $y_{vis}$ in $\protect\nu _{\protect\mu }$-mode
WSM's for data(solid), background (dashed), and background plus fitted NC
signal(dotted).}
\label{fig:yvis}
\end{figure}

Binned maximum likelihood fits are performed to the measured neutrino mode
$y_{vis}$ distribution using a model consisting of all conventional WSM
sources described and a possible $c\bar{c}$ signal. The fitter varies the NC
charm contribution in shape and level by allowing the charm mass parameter, $%
m_{c}$, to float; it also varies the normalization of the beam impurities. \
Figure \ref{fig:yvis} shows the $y_{vis}$ distribution for the data with the
background plus fitted NC charm signal superposed. The shape indicates a
preference towards including the NC charm signal, and the fit yields $%
m_{c}=1.42_{-0.34}^{+0.77}$ GeV/c$^{2}$ with a beam normalization of $1.00\pm
0.06$, where the errors are purely statistical. The fitted value of the beam
normalization validates the beam impurities model and the 
$\bar{\nu}_{\mu }$ WSM tuning procedure described in the Appendix.

In Fig. \ref{fig:yvish} the $y_{vis}$ distribution of WSM's is shown with
the additional requirement that $E_{had}$ be larger than 50 GeV. This cut
removes $85\%$ of the beam impurities while keeping 75\% of the NC charm
events. Performing the fit again on this reduced sample with the background
normalization fixed at $1.0$ yields a consistent value for the charm mass of 
$m_{c}=1.40_{-0.36}^{+0.83}$ GeV/c$^{2}$, with the error again purely
statistical.

\subsection{Systematic Errors}

Estimates of the systematic uncertainty are obtained by varying parameters
associated with background and signal simulations and event selection
criteria within known bounds. Systematic errors are assumed to be
independent and thus can be added in quadrature. More details on systematic
error studies may be found in Refs. \ref{drew fcnc prd} and \ref{drew thesis}%
. The most important contributions are from modelling the level of CC charm
events which reconstruct as WSM, from energy and momentum calibration
uncertainties, from charmed quark fragmentation, from the choice of the
gluon PDF used in the $c\bar{c}$ production model, from possible nuclear
effects, and from the beam impurity model.

The number of dimuon events which reconstruct as a WSM can be normalized
either from the total right sign muon sample and the measured charm production
fraction or from the observed number of events reconstructed with two events
in the toroid. These two normalizations disagree by $3\%$, and switching
to the latter normalization shifts $m_{c}$ by $+0.10$ GeV/c$^{2}$.
Replacing the drift-chamber-tracking-based method to reject events with
two muons by a calorimeter-pulse-height-based algorithm 
 leads to a further
shift $\delta m_{c}=+0.04$ GeV/c$^{2}$.

Shifting $E_{\mu }$ by $\pm 1\%$ calibration uncertainty changes $m_{c}$ by $%
_{-0.02}^{+0.05}$ GeV/c$^{2}$. Shifting $E_{had}$ by $\pm 0.5\%$ changes $%
m_{c}$ by $\pm 0.01$ GeV/c$^{2}$. A total systematic error of $0.05$ 
GeV/c$^{2}$ is thus attributed to calibration.

The fragmentation model, based on CC charm analysis of the same experiment 
\cite{bib:max}, is tested by using events with the same production
kinematics and the Lund string fragmentation model\cite{bib:lund}. This
change increases $m_{c}$ by $0.14$ GeV/c$^{2}$.

The only possible relatively-unknown input to the boson-gluon fusion cross
section model besides the charm quark mass is the gluon PDF. Changing
from the GRV94HO set to CTEQ4M raises $m_{c}$ by $0.04$ GeV/c$^{2}$. Varying
the CTEQ gluon PDF according to the prescription given by its authors\cite
{bib:Huston} produced a maximum variation in $m_{c}$ of $-0.04$ GeV/c$^{2}$,
and this is used as the systematic error on the gluon PDF.

It is unclear whether the EMC correction\cite{bib:emccor} for nuclear
effects should be applied to boson-gluon fusion processes, so the NC charm
fit is performed with and without it. Applying the EMC correction increases
the measured $m_{c}$ by $0.12$ GeV/c$^{2}$, and this shift is included as a
possible systematic error.

The final systematic error is due to the size of the beam impurities. The
beam fit described in the Appendix returns a normalization value and error
for each of five separate beam sub-sources. To examine the sensitivity to
each individual source, each source normalization is fixed one sigma
high and low of its best fit value, and the other sources' normalizations are
extracted. These alternative settings are then applied to $\nu $-mode beam
impurities, and $m_{c}$ is re-extracted. The largest change occurs for
scraping and the second largest for beam-produced charm. The sum, in
quadrature, of all changes is a shift in $m_{c}$ of $0.13$ GeV/c$^{2}$.

The sum of all systematic errors in quadrature is $0.26$ GeV/c$^{2}$.

\section{Summary and Conclusions}

Wrong sign muon data in $\nu _{\mu }$Fe scattering data show clear evidence
for NC open charm production. The result of a boson-gluon fusion fit with
the GRV94HO gluon PDF\ set to the data yields $m_{c}=1.40_{-0.36}^{+0.83}\pm
0.26$ GeV/c$^{2}$. This value of charm mass corresponds to a production
cross section $\sigma \left( \nu _{\mu }N\rightarrow \nu _{\mu }c\bar{c}%
X\right) =(0.21_{-0.15}^{+0.18})$ fb at an average neutrino energy $%
\left\langle E\right\rangle =154$ GeV. \ The $m_{c}$ governing NC neutrino
charm production is consistent with the value obtained from CC\ neutrino
charm production\cite{bib:baz}, from photo-production,
and from charmonium spectroscopy\cite{charmonium}.

\begin{figure}[tbp]
  \centerline{\psfig{figure=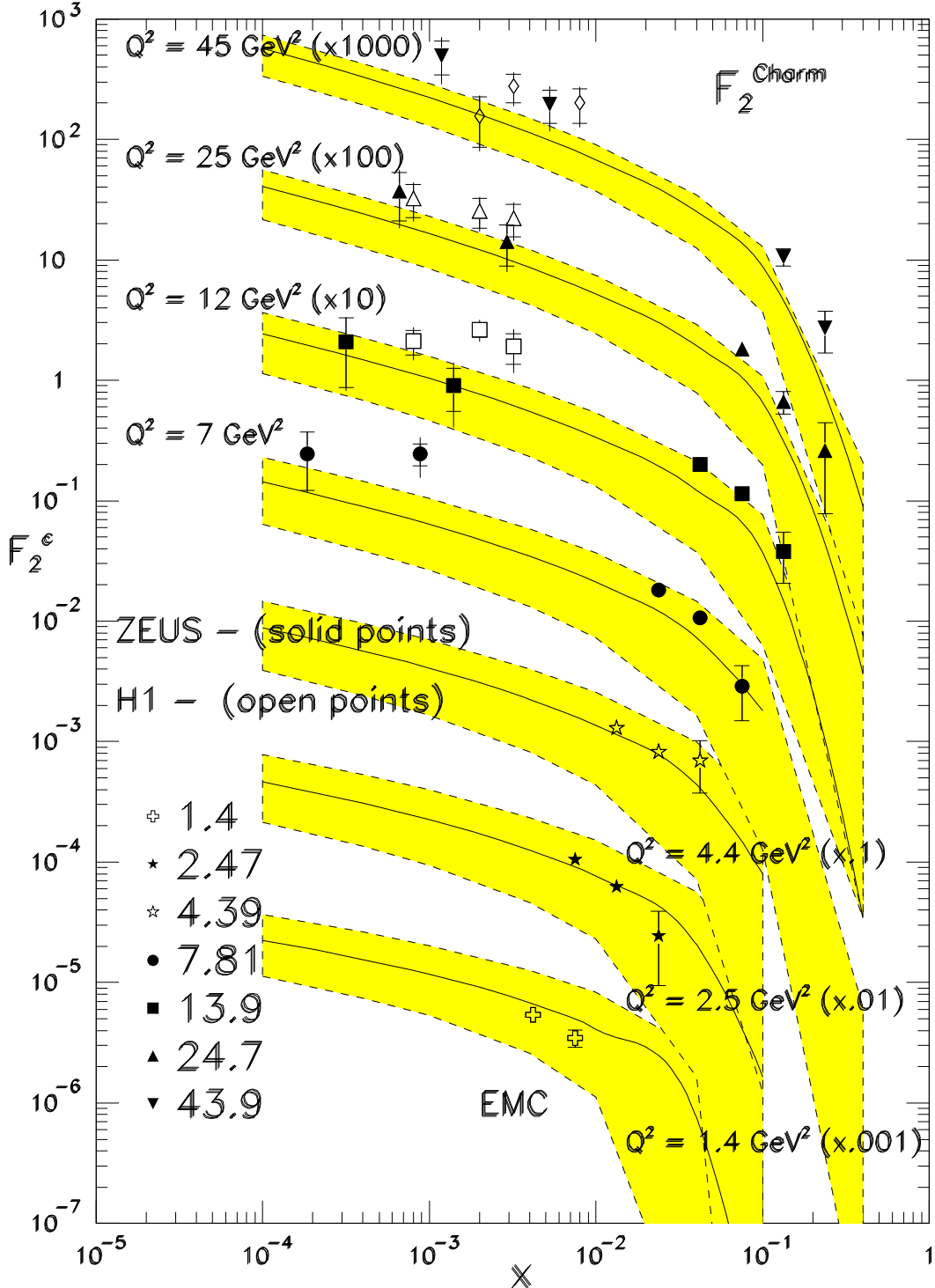,width=14cm}}
\caption{$F_{2}^{charm}$ as a function of $x$, for various $Q^{2}$. The
bands are the gluon-boson fusion cross section using $%
m_{c}=1.40_{-0.36}^{+0.83}$ and the GRV94HO gluon PDF. Data points are from
charged lepton scattering from refs. \ref{emcdata} and \ref{heradata}. 
Our data are sensitive in a region that overlaps
EMC but extends to slightly higher $Q^{2}$ and slightly lower $x$.}
\label{fig:mccros}
\end{figure}

Differential cross sections computed with this charm mass, the GRV94HO gluon
PDF, and the gluon fusion model are compared to electro-production data in
Fig. \ref{fig:mccros}. Our data is sensitive in a region that overlaps the
EMC experiment\cite{emcdata}, and it extends to slightly higher $Q^{2}$ and
slightly lower $x.$ \ It is consistent with this electro-production data
within rather large errors, providing evidence that the boson-gluon fusion
process is probe-independent, as expected from QCD.

Finally, since the NC\ charm signal can be adequately described by the
boson-gluon fusion diagram, there is no evidence for the existence of either
a perturbative or non-perturbative intrinsic charm sea from NuTeV\ data.

\acknowledgements
We would like to thank the staffs of the Fermilab Particle Physics and Beams
Divisions for their contributions to the construction and operation of the
NuTeV\ beamlines. We would also like to thank the staffs of our home
institutions for their help throughout the running and analysis of NuTeV.
This work has been supported by the U.S. Department of Energy and the
National Science Foundation.

\appendix

\section{Determination of Charm Production in $pN$ Scattering at $800$ GeV
From $\bar{\protect\nu}_{\protect\mu }$ Mode WSM Data}

Beam impurities are responsible for over $80\%$ of the WSM's in $\bar{\nu}$%
-mode. The beam impurities are due to scraping, hadrons (including charm)
that decay before the sign-selecting dipoles in the SSQT, neutral kaon
decays, muon decays and $K \rightarrow \pi \rightarrow \mu$. 
Several of these sources are not well-constrained by
previous measurements; and $\bar{\nu}$-mode WSM's can be used to improve
knowledge of their normalization. For the case of WSM's from charm decay, this
effectively amounts to performing a new measurement of $\sigma \left(
pN\rightarrow c\bar{c}X\right) $.

Secondary $\pi $'s and $K$'s are modelled by Malensek's parameterization\cite
{bib:mal} of Atherton's data\cite{bib:ath}. The interaction of secondaries
with beam elements (scraping) is modeled by GHEISHA\cite{bib:ghes}. The $%
K^{0}$ production is handled by extending Malensek's charged kaon
parameterizations using the quark counting relation $K_{L}^{0}=\left(
3K^{-}+K^{+}\right) /4$.  The $K \rightarrow \pi \rightarrow \mu$ process
is correctly modeled.
Muon decay is a well understood process;  however,
there is some uncertainty in the polarization ${\cal P}$ of the beam which
should lie within the range ${\cal P=}0.1_{-0.0}^{+0.2}$. Two experiments
measure the inclusive cross section for production of $D^{\pm }$ and $D^{0}/%
\overline{D^{0}}$ mesons with an 800 GeV proton beam(Ammar {\it et al.}\cite
{bib:amm} and Kodama {\it et al.}\cite{bib:kod}). The weighted average of
their production parameters are used as the starting value in this analysis.

\begin{figure}[hbt]
 \psfig{figure=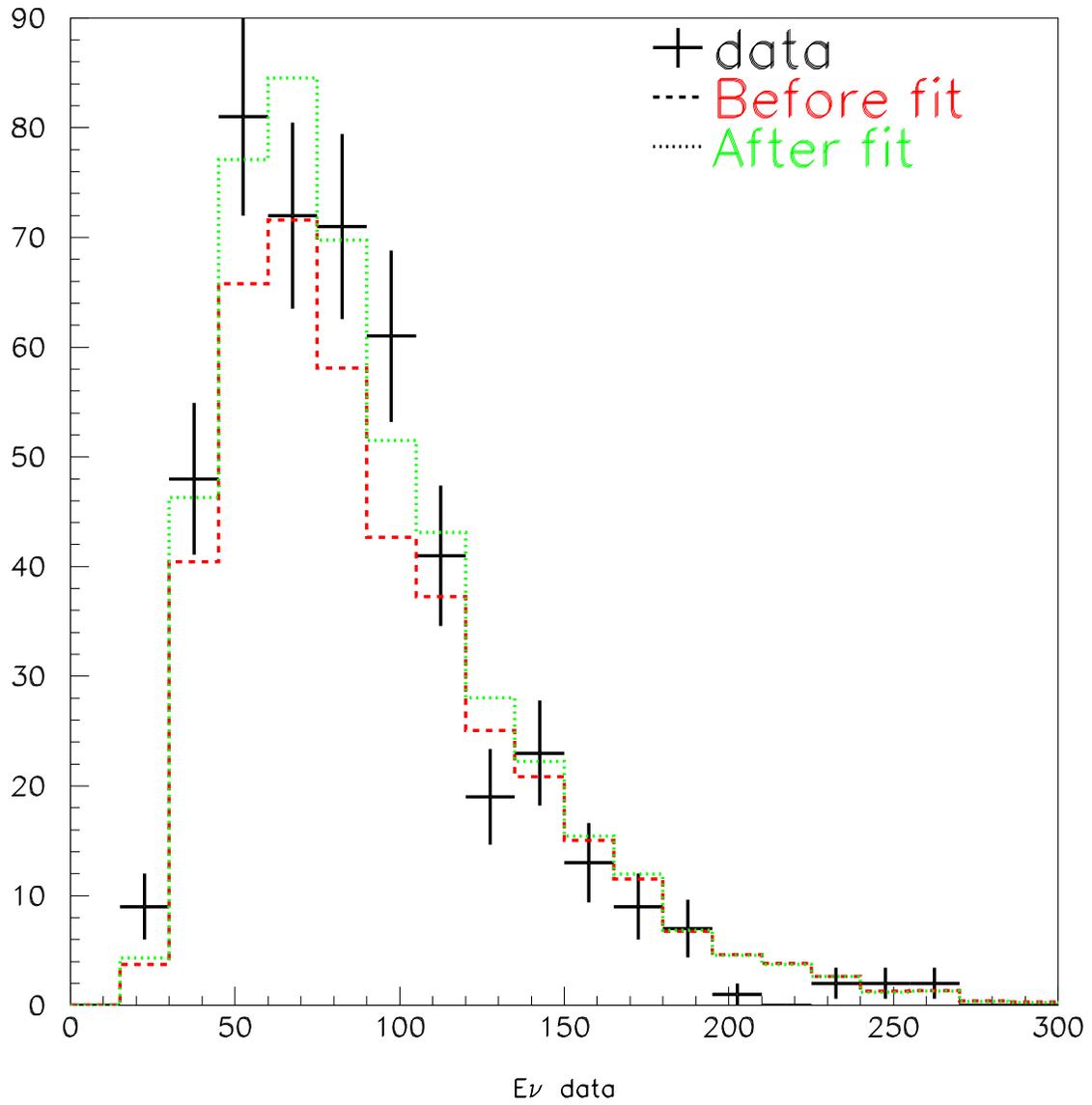,width=17.cm} 
\caption{The $E_{\nu}$ distribution of WSM events in 
$\bar \nu$-mode before and after the fit.}
\label{fig:fitpar}
\end{figure}

A model of the WSM's is constructed from the beam sources added, 
with adjustable
weights, to non-beam sources (CC and NC charm, $\pi /K$ decay in the shower,
and charged mis-measurement), and a binned likelihood fit is performed
jointly to the neutrino energy distribution and the vertical position of WSM
neutrino interactions in $\bar{\nu}_{\mu }$ mode . The fit constrains the
weight of beam sources, other than charm production, to be consistent with
1.0 within their estimated {\it a priori} errors. Table \ref{tab:bres} lists
these errors and gives the fit results.  The only significant deviations
from unity of any source normalizations are in the scraping and charm
contributions. The charm result indicates that $\sum_{i}\sigma (D_{i})\times
BF(D_{i}\rightarrow \nu _{\mu })=(9.8\pm 2.2)\;\mu $b be increased of almost
50\% over the {\it a priori} estimate. Recalling that there are two mesons
for each $c\overline{c}$ pair, using $BF(c\rightarrow \mu )=9.9\pm 1.2\%$%
\cite{bib:baz}, and assuming linear A dependence, one obtains $
\sigma(p+N\rightarrow c\overline{c})=(49 \pm 11) \mu b$.  
Figure \ref{fig:fitpar} shows the agreement between neutino energy 
($E_{\nu}$) of data 
and MC before and after the fit.

The dominant systematic error, on $\sigma(p+N\rightarrow c\overline{c})$, 
is 5.0 $\mu $b due to the uncertainty in $%
BF(c\rightarrow \mu )$. The only other large systematic error is 2.3 $\mu $b
due to the different methods of rejecting events with two muons described
in the systematic errors above. Systematic errors due to the 
normalization of non-beam sources,
energy calibrations, parameterization of $p_{t}$ and $x_{f}$, and the use of
the $y_{vis}<0.5$ cut are small. The total of all of these sources is 5.6 $%
\mu b$ yielding the final result of $\sigma(p+N \rightarrow c\overline{c}%
)=(49\pm 11\pm 6)$ $\mu $b$.$

Using PYTHIA's\cite{bib:lund} fragmentation of c quarks into mesons one can
transform Kodama's and Ammar's measurements into the measurements of $\sigma
(p+N\rightarrow c\overline{c})$ found in Table \ref{tab:ka}. NuTeV's
measurement is consistent with these previous measurements, and has smaller
errors.

\begin{figure}[tbp]
  \centerline{\psfig{figure=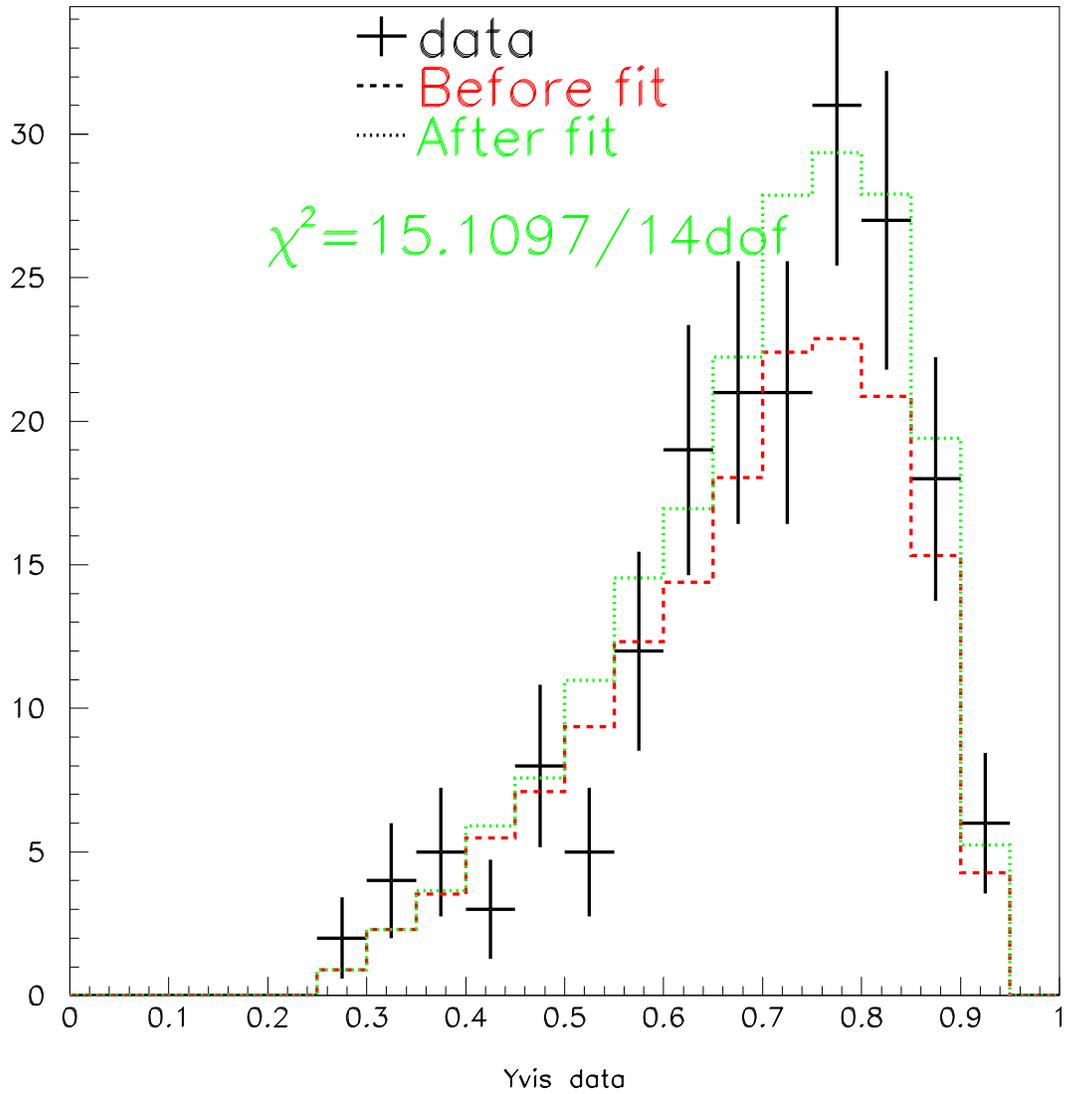,width=\textwidth}}
\caption{Distribution of $y_{vis}$ for WSM's for data(solid),
backgrounds(dashed), and background plus NC signal(dotted) with an
additional requirement $E_{had}\geq 50$ GeV.}
\label{fig:yvish}
\end{figure}

\begin{table}[h]
\begin{tabular}{ccc}
Source & $\nu $-mode$\left( \%\right) $ & $\bar{\nu}$-mode$\left( \%\right) $
\\ \hline
Beam Impurity & 67 & 83 \\ 
Charged Current Charm & 19 & 8 \\ 
Charge Misidentification & 5 & 5 \\ 
Neutral Current Charm & 5 & 2 \\ 
Neutral Current $\pi /K$ decay & 2 & 1 \\ 
Charged Current $\pi /K$ decay & 1 & 1 \\ \hline
\end{tabular}
\caption{Percentage of WSM's for each source in a given mode.}
\label{tab:fracs}
\end{table}

\begin{table}[h]
\begin{tabular}{lll}
& $\nu $-mode & $\bar{\nu}$-mode \\ 
& E$\nu >$ 20 GeV & E$\nu >$ 20 GeV \\ \hline
scraping & 53\% & 24\% \\ 
charm & 10\% & 25\% \\ 
K$^{0}$ & 12\% & 16\% \\ \hline
other prompt & 9\% & 22\% \\ 
muon decay & 11\% & 11\% \\ 
$K\rightarrow \pi \rightarrow \mu $ & 5\% & 2\% \\ \hline
\end{tabular}
\caption{The percentage of beam impurities due to a given source in each
mode.}
\label{tab:ws50}
\end{table}

\begin{table}[h]
\caption{Results of $\bar\protect\nu$-mode beam fits. 
* The {\it a priori} charm error is not used to
constrain this fit.}\label{tab:bres} \centering
\begin{tabular}{|c||c|c|c|}
\hline
Source & Value & Error & {\it A priori} Error Estimate \\ \hline
charm & 1.47 & 0.33 & 0.30* \\ 
$K^0$ & 1.01 & 0.29 & 0.20 \\ 
scrape & 1.22 & 0.34 & 0.40 \\ 
other & 1.00 & fixed & 0.03 \\ 
muon & 0.95 & 0.11 & 0.07 \\ 
prompt & 1.02 & 0.21 & 0.10 \\ \hline
\end{tabular}
\end{table}
\begin{table}[h]
\caption{Previous charm meson production cross-sections transformed into
charm quark production cross-sections. }\label{tab:ka} \centering
\begin{tabular}{|l|c|c|}
\hline
Exp & $\sigma(p+p \rightarrow c\overline{c})$ from $D^{\pm}$ measurement & 
$\sigma(p+p \rightarrow c \overline{c})$ from $D^{0}$ measurement \\ \hline
Kodama(1991) & 75$\pm$18$\pm$28 & 47$\pm$4$\pm$16 \\ 
Ammar(1988) & 51$\pm$8$\pm$13 & 27$^{+11}_{-9}\pm$7 \\ \hline
\end{tabular}
\end{table}

\end{document}